# Coherent Epitaxial Semiconductor - Ferromagnetic Insulator InAs/EuS Interfaces: Band Alignment and Magnetic Structure

Yu Liu[1], Alessandra Luchini[2], Sara Martí-Sánchez[3], Christian Koch[3], Sergej Schuwalow[1], Sabbir A. Khan[1], Tomaš Stankevič[1], Sonia Francoual[4], Jose R. L. Mardegan[4], Jonas A. Krieger[5], Vladimir N. Strocov[5], Jochen Stahn[5], Carlos A. F. Vaz[5], Mahesh Ramakrishnan[5], Urs Staub[5], Kim Lefmann[2], Gabriel Aeppli[5,6], Jordi Arbiol[3,7], Peter Krogstrup[1,*]

[1]Center for Quantum Devices, Niels Bohr Institute, University of Copenhagen & Microsoft Quantum Materials Lab Copenhagen, Denmark.

[2]Niels Bohr Institute, University of Copenhagen, DK-2100 Copenhagen, Denmark.

[3]Catalan Institute of Nanoscience and Nanotechnology (ICN2), CSIC and BIST, Campus UAB, Bellaterra, 08193 Barcelona, Catalonia, Spain.

[4]Deutsches Elektronen-Synchrotron DESY, Hamburg 22603, Germany.

[5]Paul Scherrer Institute, CH-5232 Villigen, Switzerland.

[6]ETH CH- 8093 Zürich and EPFL CH-1015 Lausanne, Switzerland.

[7]ICREA, Pg. Lluís Companys 23, 08010 Barcelona, Catalonia, Spain.

**ABSTRACT: Hybrid semiconductor - ferromagnetic insulator heterostructures are interesting due to their tunable electronic transport, self-sustained stray field and local proximitized magnetic exchange. In this work, we present lattice matched hybrid epitaxy of semiconductor - ferromagnetic insulator InAs/EuS heterostructures and analyze the atomic-scale structure as well as their electronic and magnetic characteristics. The Fermi level at the InAs/EuS interface is found to be close to the InAs conduction band and in the bandgap of EuS, thus preserving the semiconducting properties. Both neutron and X-ray reflectivity measurements show that the ferromagnetic component is mainly localized in the EuS thin film with a suppression of the Eu moment in the EuS layer nearest the InAs. Induced moments in the adjacent InAs layers were not detected although our ab initio calculations indicate a small exchange field in the InAs layer. This work presents a step towards realizing high quality semiconductor - ferromagnetic insulator hybrids, which is a critical requirement for development of various quantum and spintronic applications without external magnetic fields.**

## 1. INTRODUCTION

Lifting electron spin degeneracy in semiconductor materials plays a central role in spintronics and emerging quantum technologies. An interesting path is to combine semiconductors (SE) and ferromagnetic insulators (FMI) as the carrier quality and tunability of the semiconductor potentially could remain intact while the spin degeneracy is lifted by a ferromagnetic exchange coupling. But the material requirements are many folds and depend on several details such as energy band alignment and structural quality which limit the choice of materials. Semiconductors with narrow direct bandgaps such as InAs and InSb, which have high electron mobilities and strong spin-orbit coupling, are interesting for a broad range of applications such as infrared detectors, terahertz radiation sources and quantum applications. For example, hybrid III-V semiconductor - superconductor (SE/SU) nanowire materials[1-6] has potential as a fault tolerant basis for quantum information processing[7-10]. However, large external magnetic fields are needed to enter the topological regime which weaken the superconducting paring and complicate scaling to technologies. Therefore, it is of interest to integrate materials that are intrinsically topological, ideally without the need for external applied fields. Composite tri-crystals using FMIs in close proximity to the SE/SU structure have been proposed as a solution to realize the topological phase[11], where the effective Zeeman splitting is induced by exchange coupling with the FMI.

Recently, magnetic exchange coupling with Zeeman splitting energy larger than 2.0 meV has been reported on graphene when combined with EuS[12]. Magnetization up to 1.84 $\mu_B$/Bi can go into $Bi_2Se_3$ due to the proximity to EuS[13]. Spin-polarized tunneling has been reported on Au/EuS/Al[14, 15] and extra Zeeman splitting in the superconducting quasiparticle density of states has been observed in EuS/Al films[16]. EuS has a cubic rock salt structure with a small bulk lattice mismatch to the InAs zinc blende phase (1%, <001> vs. <001>), and, importantly, single crystal epitaxy can be obtained at temperatures that are compatible with InAs stability.[17] Therefore, for fabricating InAs/Al/FMI systems, Eu-based chalcogenides[18, 19] have larger potential than other FMIs reported in literature, such as ferrites[20, 21], garnets[22] and layered Cr-based trihalides[23, 24]. On the other hand, it is not always feasible to induce the exchange field by proximity. It is well-known that, the spontaneous spin coupling will favor ferromagnetic order if complying with the Stoner criterion[25, 26], that is, its exchange energy multiplied by the density of states is larger than unity. Several previous works have reported the absence of proximity effects in normal metal – FMI systems[27-31]. Therefore, it would also be necessary to figure out whether there are proximity effects in InAs-FMI hybrid structures as a first step towards the development on complex SE/SU/FMI tri-crystal systems. Here, we investigate the bi-crystal epitaxy of InAs/EuS 'SE/FMI' system, which shows itself as a unique hybrid material, with well-defined epitaxial matching, semiconducting band alignment and localized magnetic structure.

## 2. EXPERIMENTAL SECTION

Planar InAs layers of 50 nm were grown on 2-inch undoped (001) zinc blende GaSb wafers (Wafer Technology Ltd.) in a solid-source Varian GEN-II MBE system. The natural oxidized layers were desorbed from the InAs by heating the substrate to a temperature of 525 °C (measured with a pyrometer) under $As_2$ overpressure protection for 360 s. The InAs layers were grown at a substrate temperature of 500 °C (measured with a pyrometer) and the growth rate was ~0.1 nm/s. The substrates were cooled to 300 °C under $As_2$ overpressure and further cooled to 200 °C after closing for the As flux. After the reaching background pressure below $10^{-9}$ Torr the samples were transferred to the connected UHV deposition chamber with EuS (without breaking UHV) which were grown with electron beam evaporation[13]. Growth procedures were not initiated before both chambers reached background pressures below $10^{-10}$ Torr. The substrates were placed about 45° tilt towards the EuS source and the EuS growth rate was calibrated with combined TEM and quartz crystal readout. The substrate temperature during growth was 180 °C (measured with a thermo-coupling back sensor) and the average growth rate is 0.02 nm/s. Amorphous As or AlOx was deposited to protect the surface from oxidation. The samples were structurally characterized by High Angle Annular Dark Field (HAADF) imaging using an aberration corrected scanning transmission electron microscope (AC-STEM) Titan FEI Microscope. The cross-section lamellae were obtained using a Focused Ion Beam HELIOS 600 FIB system. The energy band alignment was measured using soft-X-ray angle resolved photoemission spectroscopy (SX-ARPES) with synchrotron radiation excitation at the Swiss Light Source, Paul Scherrer Institute, Switzerland, using the SX-ARPES end station of the high-resolution undulator beamline ADvanced REsonant Spectroscopies (ADRESS) operating in the energy range from 300 eV to 1.6 keV.[32, 33] The combined beamline and analyzer resolutions were better than 170 meV and 85 meV at hυ = 1025 eV and hυ = 405 eV, respectively. The sample temperature was ~20 K, above the Curie temperature of EuS. The Fermi level reference was set by the SX-ARPES spectrum of a gold contact on the hybrid chip.

For magnetic profile studies, polarized neutron reflectivity (PNR) measurements were performed at the reflectometer AMOR at PSI. Amor is operated in the focused beam (SELENE) mode[34, 35] and time-of-flight (TOF) mode and it has a vertical scattering plane. PNR data were collected under a 0.1 T magnetic field parallel to the sample surface at temperature above (50 K) and below (2 K) the EuS $T_C$ (~16 K). X-ray absorption near edge structure (XANES) and X-ray magnetic circular dichroism (XMCD) measurements were carried out at beamline P09 at PETRA III (DESY) in a fluorescence mode[36]. In order to enhance the signal and decrease the air absorption, a silicon drift detector was placed in vacuum and approximately 150 mm from the thin films. The thin films were mounted inside a displex cryostat with a base temperature of 5 K. XMCD spectra were acquired via a fast helicity switching mode at 11.5 Hz in which the left- and right-handed circularly polarized X-rays were obtained using diamond phase plates[37]. The resonant x-ray reflectivity (RXRR) measurements were performed at the RESOXS[38] station of the SIM beamline[39] at the Swiss Light Source of the Paul Scherrer Institute.

## 3. RESULTS AND DISCUSSION

**Epitaxy of EuS on InAs.**

The stable crystal structure of EuS is known as cubic rock-salt (space group Fm-3m) with lattice constant of 5.969 Å (ICSD 30206). Therefore, the lattice mismatch between the two cubic crystals, rock salt EuS and zinc blende InAs (ICSD 24518, space group F-43m, lattice constant 6.048 Å), is only ~1%. As we show below in Fig. 1a, such a small mismatch allows the epitaxy of elastically strained and fully coherent InAs/EuS hybrid crystals. The cross-section lamellae were obtained perpendicular to the growth direction of the selected areas where the zone axis (observed direction) is along [110] of zinc blende InAs and rock salt EuS structures, with an interface (001) normal (for details, see Fig. S1 of indexed power spectrum in supplementary information). Using geometric phase analysis (GPA), the displacement fields of the atomic columns are obtained to provide quantitative strain maps from standard atomic resolution HAADF STEM images[40, 41]. In Fig. 1b, we show a GPA dilatation map of the out-of-plane (002) planes, corresponding to the STEM image seen in Fig. 1a. The dark blue region of about 2 monolayers shows a lattice compression of remarkable 10%, while the in-plane dilatation map of (2-20) planes in Fig. 1c demonstrates a matched interface without misfit dislocations. The rotation maps on (002) and (2-20) in supplementary information, Fig. S2, confirm the coherence of the interface. However, interfacial displacement appears in the (1-1-1) and (-11-1) rotation maps due to non-orthogonal strain components at the interface. To explore the atomic positions, the enlarged HAADF STEM image of the interface is shown in Fig. 1d. The atom species are analyzed based on the chemical sensitivity (Z-contrast) of the HAADF STEM[42, 43], where the intensity of the scattered atoms in HAADF is proportional to $Z^\zeta$ with $\zeta$ ranging from 1.2 to 1.8. We verified the intensities of the atomic columns in InAs and EuS bulk all have the predicted HAADF STEM intensities except the two monolayers at the interfaces which show a significant decrease in intensity. This is because the average column density (i.e. the combined mass measured in line of sight) is lower due to single layer intermixing causing a displacement from the ideal atomic rows as apparent from the InAs and EuS phases. This hypothesis is supported by HAADF intensities simulated with the Rhodius software[44, 45], which present an interface including atomic steps in the lateral viewing direction as sketched in Fig. 1e (see details in Fig. S3 in supplementary information). By comparing the intensity distribution in Fig. 1d with the simulations presented in Fig. S3, we estimate the atomic configuration at the interface as shown in Fig. 1f. Based on this analysis, a topview model of the interface is shown in Fig. S4. The order of the plane rotational symmetry (PRS) of EuS on InAs is 2 and its bi-crystal variant is 1 (see ref. 2 for analysis), consistent with the single crystalline structure observed across the wafer.

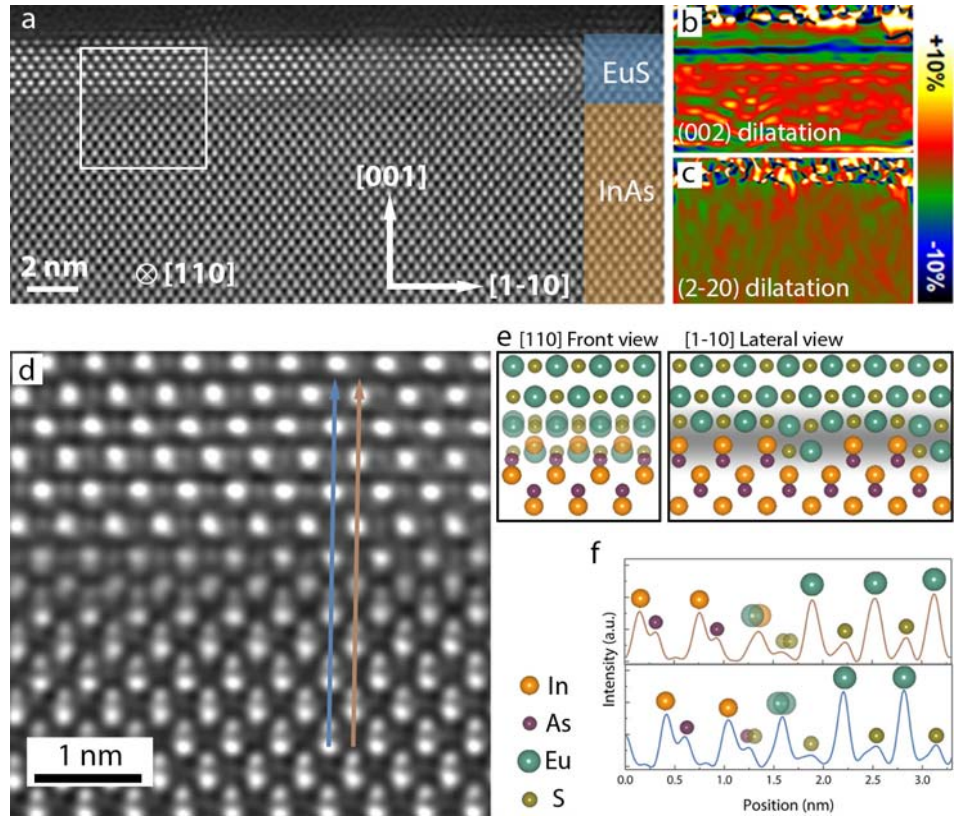

**Figure 1. Cube-on-cube epitaxy of EuS on InAs. (a)** HAADF-STEM micrograph of a planar epitaxial InAs/EuS film. The white box defines the enlarged region shown in **(d)**. **(b-c)** The GPA dilatation map of the out-of-plane (002) planes shows a single interface layer with ~10% compression, while the in-plane (2-20) dilatation shows a dislocation free and fully coherent epitaxial match. **(d)** The atom-resolved HAADF AC-STEM image to show atomic arrangement near the interface. **(e)** The proposed model of the InAs/EuS interface with zone axis [110] and [1-10] based on the HAADF intensity image simulation. Structural modeling of the interfaces was performed using VESTA[46]. More details are in Fig. S3. **(f)** The HAADF intensity as a function of positions using the image from **(d)**. The arrows with the corresponding colors in **(d)** show where the HAADF STEM intensity profiles are extracted from. The atom species are determined by HAADF STEM intensity simulation. Note that the interface is just a projection of [1-10] from a lateral side, so there is no overlapping.

**Band alignment at the InAs/EuS interface.**

The energy band alignment at SE/FMI interfaces will play an important role on the detailed proximity effects between the two types of materials. Moreover, EuS has been reported to be conducting due to doping and impurities.[47, 48] Therefore, it is necessary to study the details of the quality and interfacial band structure in these samples. With SX-ARPES (see details in experimental section) the band structure in the first Brillouin zone of InAs is measured along the Γ-X direction for a bare InAs sample at photon energy 405 eV as shown in Fig. 2a. The corresponding SX-ARPES data for an epitaxially grown planar InAs/EuS interface in Fig. 2b (photon energy increased to 1025 eV in order to increase the probing depth towards InAs) shows a strong intensity enhancement between -2.5 ~ -1.0 eV coming from the EuS 4f electron states. These states are located below the maximum of valence bands in InAs, which is emphasized by subtracting the non-dispersive (angle-integrated) spectral component as shown in Fig. 2c. It is known that the gap between the bottom of the conduction bands and the 4f states of EuS is ~1.7 eV[49], which is much wider than the bandgap of InAs, 0.4 eV. Therefore, EuS, as an insulator, has no contribution to the conductivity of the InAs/EuS interface. It should be noted that the InAs surface typically hosts a surface charge, creating band bending which manifests itself by a tiny pocket of conduction-band derived quantum well states observed in Fig. 2a at the Γ-point. The SX-ARPES results on interfacial InAs/EuS band alignment (Figs. 2b-2c) show the position of the Fermi level close to the InAs conduction band in a manner similar to pristine InAs, lying in the bandgap of EuS. This allows formation of an emergent two-dimensional electron gas in InAs in the vicinity of the interface, necessary for field-effect devices, without opening additional conduction channels in the hybrid system.

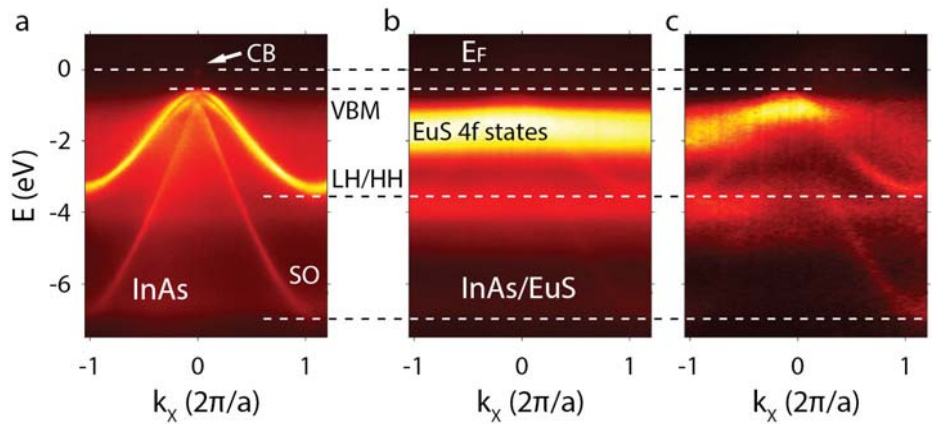

**Figure 2. SX-ARPES of InAs/EuS hybrid structure. (a)** SX-ARPES spectra of planar InAs (001) along the Γ-X (in-plane) direction of the first Brillouin zone. The Fermi level ($E_f$, set to zero), CB (conduction band), VBM

(valence band maximum), LH/HH (light hole / heavy hole band) and SO (split-off band) of InAs are marked. The conduction-band derived quantum-well states due to the surface band bending are seen. **(b)** Corresponding spectra of an epitaxially grown planar InAs/EuS heterostructure along the same direction as in **(a)**. 4f states of EuS are marked. No EuS states are observed at the Fermi level. **(c)** Subtracting non-dispersive spectral component from the spectra in **(b)** emphasizes that the EuS 4f electron energy level is below the VBM in InAs and that interfacial quantum-well states form in InAs.

**Distribution of magnetic moments at the interface.**

To study the magnetic structure and the potential exchange field at the InAs/EuS interface, we employed three complimentary characterization methods, PNR, XMCD and RXRR measurements.

During a PNR experiment[50], the intensity of neutrons with spins polarized either parallel (up or +) or antiparallel (down or -) to the external magnetic field was measured as a function of the scattering vector $q = \frac{4\pi}{\lambda}\sin(\theta)$, where θ is the angle between the incident neutron beam and the sample surface and λ is the neutron wavelength. The reflectivity is calculated for both spin polarization states as the ratio between the reflected and the incoming intensities: $R^+ = \frac{I_r^+}{I_0^+}$ and $R^- = \frac{I_r^-}{I_0^-}$. If the sample has no magnetic structure, $R^+$ and $R^-$ will exhibit the same q-dependence, which depend only on the density depth profile of the sample. If the sample is magnetic, the neutrons will interact with both the magnetic moments and the nuclei of the surface layers, resulting in different reflectivity profiles according to their spin-polarization directions.

The experimental data were analyzed with MOTOFIT[51] where the sample is considered to be composed of a stack of layers, each of which is defined by a scattering length density (SLD or ρ), a thickness and a surface roughness. The SLD describes the interaction between the neutrons and the nuclei in the sample and is considered as the sum of nuclear scattering length density (NSLD or $\rho_n$), which takes into account the coherent scattering cross section, absorption scattering length density (ASLD or $\rho_a$), which considers the possibility for nuclei to absorb neutrons, and magnetic scattering length density (MSLD or $\rho_m$), which accounts for the contribution brought by magnetization. Figure 3a shows the PNR data collected at 2 K and 50 K with an external magnetic field corresponding to 0.1 T. Data collected at two different temperatures were simultaneously analyzed with the same structural model, but only the MSLD contribution is taken into account for the data collected at 2 K (the optimized structural parameters are reported in Table S1 in supplementary information). In principle, PNR data can be used to extract the thicknesses and densities of the layers. However, the negligible difference in the real part of the nuclear scattering length density ($\Delta$NSLD = 0.33 × 10$^{-4}$ nm$^{-2}$) between InAs (NSLD$_{InAs}$ = 1.89 × 10$^{-4}$ nm$^{-2}$) and EuS (NSLD$_{EuS}$ = 1.56 × 10$^{-4}$ nm$^{-2}$) limits the accuracy in the determination of the EuS layer thickness. Nevertheless, the EuS thickness is estimated to be (2.0 ± 0.2) nm, which is close to the value obtained using STEM. On the other hand, the thickness of the magnetic layer approximately matches the value for EuS based on the MSLD profile in Fig. S5. With the experimental resolution (~ 2 Å considering the roughness is 3 Å), we cannot conclude whether or not there is a weak net magnetization within InAs.

XMCD is a powerful tool to understand the element specific origin and/or distribution of magnetization in a material[52, 53]. XANES and XMCD spectra were collected at different temperatures, with the beam energy tuned to the Eu $L_3$ (6977 eV), In $L_3$ (3730 eV), and at the As K (11867 eV) absorption edges. The XANES signal was recorded averaging the absorption spectra obtained with left and right circular polarization, while the XMCD signal was obtained from the difference. An assembly of permanent magnets mounted on the sample holder provided an external magnetic field of 0.1 T along the incident beam wave vector **k**. Figure 3b shows the typical XANES and XMCD at the Eu $L_3$-absorption edge in which the empty 5d orbitals are being probed via a dipolar transition ($2p_{3/2} \rightarrow 5d$). The top panel in Fig. 3b displays an intense single white line peak located at 6977 eV, which is characteristic of magnetic Eu ions with valence state 2+, i.e., S-state (S = 7/2, J = 7/2). We emphasize that the typically observed contributions from non-magnetic $Eu^{3+}$ ions (S = L = 3, J = 0), which is a measure of derivation from stoichiometric composition, should appear approximately 6-8 eV above the absorption edge. This is completely absent from the data and a signature of a high crystalline quality[54]. In order to investigate the ferromagnetic moment of the Eu ions, XMCD measurements were performed at three different temperatures. As shown in the bottom panel of Fig. 3b, a strong magnetic signal is observed at 5 K which indicates a strong spin polarized 5d orbitals by the $4f^7$ states. Upon warming up the system towards the critical (Curie) temperature, the thermal energy destroyed the long-range magnetic order and therefore a strong suppression of the magnetic signal was observed at 16 K which completely vanished at 23 K.

For the RXRR measurements we measured the Eu $M_5$-absorption edge near 1131.5 eV. The magnetic moment profile of EuS includes a ~1.4 Å thick InAs/EuS interfacial layer with a reduced moment by an effective factor of 0.38 compared to the bulk EuS layers (see supplementary Fig. S7). Intermixing at the interface, as described in Fig. 1, will dilute the Eu moment density as well as potentially create regions where the antiferromagnetic tendency of EuAs bonds could manifest itself. The reduced number of nearest neighbors (relative to that of next nearest neighbors) will also serve to enhance the importance of the next nearest neighbor antiferromagnetic interactions in EuS similar to what occurs in $Eu_xSr_{1-x}S$[55], which will further reduce the ferromagnetic moment in the Eu rich side of the interface.

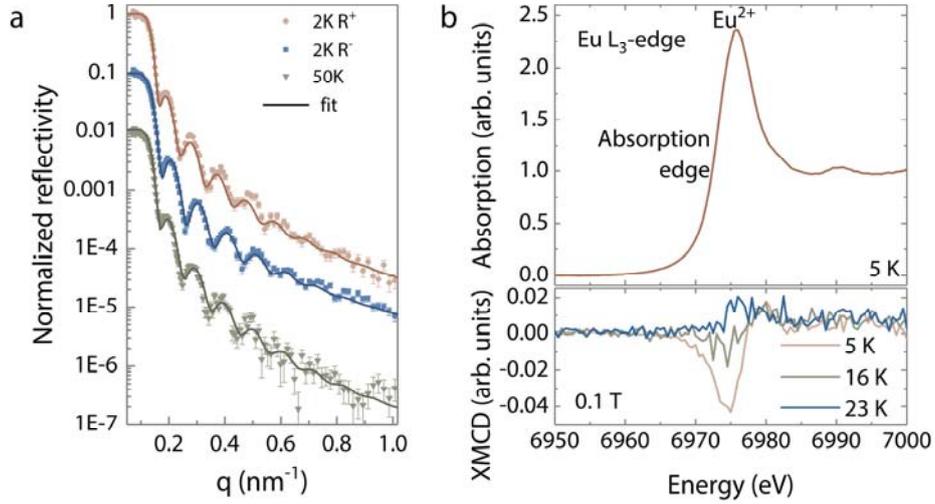

**Figure 3. PNR and XANES/XMCD of the InAs/EuS hybrid structure. (a)** Experimental data of PNR collected at 50 K and 2 K together with the respective fitting curves. The reflectivity from the sample was measured as a function of the scattering vector q. For the data measured at 2 K, we show $R^+(q)$ and $R^-(q)$ for neutrons polarized parallel or antiparallel to the external magnetic field (0.1 T). The difference between $R^+(q)$ and $R^-(q)$ arises from the magnetization in the sample. **(b)** Top panel shows the XANES spectrum collected at 5 K with the X-ray energy beam tuned to the Eu $L_3$-edge; Bottom panel shows the XMCD signal collected at the aforementioned edge under a magnetic field of 0.1 T and measured below (5 K), around (16 K) and above (23 K) the EuS Curie temperature.

Based on the results of Eu, we further carried out XANES and XMCD measurements at the In $L_3$-edge, which should be feasible even though In is not a magnetic element. In previous reports magnetic moments have for example been found to originate from the $p_z$ orbitals of carbon atoms around divacancies in noble gas implanted SiC[56, 57], as well as ferromagnetism in Si which arises from 3s*3p* hybridized states of V6 (hexagonal ring constituted by six silicon single vacancies)[58] has been demonstrated after neutron irradiation. Moreover, magnetic moments from d states are found to be localized around Ta atoms in the Weyl semimetal TaAs after proton irradiation[59]. The discovery of d-orbital magnetization at the interface between non-magnetic oxides also shows the high sensitivity of such technique[60, 61]. The distinct absorption edge at the In $L_3$-edge is detected at 3731 eV and a broad feature in the XMCD signal is seen at 5 K as shown in Fig. S8. However, the XMCD signal remains the same at 20 K, i.e. above the Curie temperature. Similar results are found at the In $M_1$ and $M_{2,3}$-edges. Accordingly, we conclude that a magnetic moment at the In atoms smaller than our detection limit. Similarly, no significant XANES signal was detected at the As absorption

K-edge, but this is due to the small amount of As atoms probed at the larger penetration depth at 11.867 keV.

The strength and extent of the proximity effect will depend on the band alignment and wave function overlap of the semiconductor and the ferromagnetic states. There are several examples that semiconductors could be ferromagnetic relying on the wave function tails[62-65]. However, if the wave function does not extend far enough from EuS into InAs, PNR will not be able to confirm the proximity effect because of the detection limit ~ 0.1 $\mu_B$ (the magnetic moments can be detectable in a few Å, only if they are larger than 1 $\mu_B$), while the XMCD signal at the energy chosen could be too weak. To verify this speculation, we carried out first-principles calculations based on the spin-polarized density-functional theory. First-principles calculations on the basis of density functional theory with spin polarization were carried out by using the Perdew-Burke-Ernzerhof exchange-correlation functional based on the generalized gradient approximation[66]. All calculations were executed in the Cambridge Serial Total Energy Package[67]. The cut-off energy was set to 480 eV for the plane-wave basis to represent the self-consistently treated valence electrons, as ultrasoft pseudopotentials depicted the core-valence interaction[68]. Self-consistent field calculations had a tolerance of $5.0 \times 10^{-5}$ eV/atom. The conjunctions between InAs surface and EuS surface for the simulation were built according to the structural information obtained from STEM images. The supercell representing the InAs/EuS interface contains two InAs unit cells, two EuS unit cells and a 10 Å vacuum layer to isolate two components avoiding unnecessary approaching. Figure 4 shows the spin-resolved charge density isosurface of interfacial states associated with the InAs/EuS hybrid structure in a 32-atom supercell. It indicates that spins are mostly localized around Eu with a magnetic moment of ~7 $\mu_B$/atom. Only weak spin polarization (~0.08 $\mu_B$/atom) is found in the space between In atoms which are the nearest neighbors of EuS. There are also minor magnetic moments located around As atoms (~0.04 $\mu_B$/atom). The symmetric shape of spin density isosurface around Eu suggests that those spins favor strong localization probably because they arise from the localized Eu 4f states. Therefore, although there could be a static magnetic proximity effect at the InAs/EuS interface, our results suggest that it is weak compared with that reported on graphene/EuS[12] or Bi$_2$Se$_3$/EuS[13].

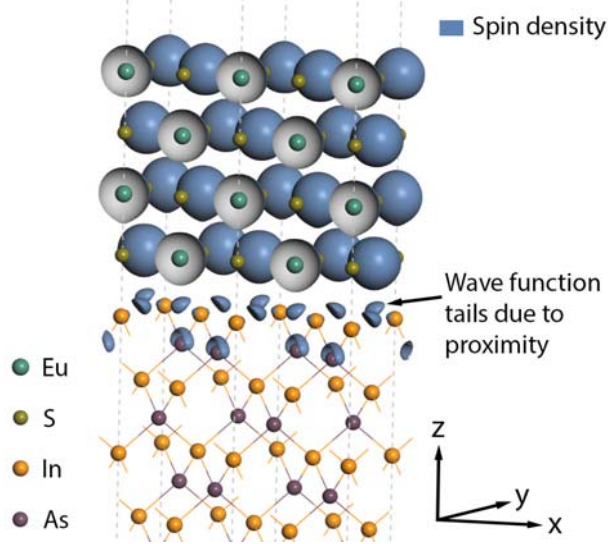

**Figure 4. Spin resolved isosurface charge density plot.** The plot is drawn in grey blue of a 32-atom supercell representing InAs/EuS interface (isovalue is 0.016 e/Å$^3$), demonstrating both the localized and the extended states due to proximity. It indicates that the magnetic moments are mostly localized around Eu. Only weak spin polarization is in the space between In atoms which are the nearest neighbors of EuS. There are minor magnetic moments located around As atoms nearest to the interface.

Based on these results it is unlikely that a strong exchange coupling leading to significant spin lifting in the semiconductor is possible with this material combination. However, for the design of zero-field topological wires, as motivated in the introduction, SE/SU/FMI tri-crystal systems are needed to enter the topological regime. Thus, EuS might induce the spin splitting in the Al[16] which is known to proximitize InAs[3-6]. Another approach could be to enhance the magnetic proximity effect on InAs by pushing the wave function into InAs side with gate tuning. Nevertheless, the hybrid epitaxial InAs/EuS with fully coherent matching and high crystalline quality is required for integration of functional hybrid quantum and spintronic applications without external magnetic fields.

## 4. CONCLUSIONS

In summary, we present hybrid epitaxy and characterization of semiconductor - ferromagnetic insulator (InAs/EuS) lattice matched interfaces. The EuS thin film on (100) zinc blende InAs is fully coherent and free from misfit dislocations. The interfacial InAs/EuS band alignment leads to a position of the Fermi level close to InAs conduction band and in the bandgap of EuS, thus preserving semiconducting tunability. PNR measurements indicate that most of magnetic moments are localized in the EuS layer and that there is no static magnetic proximity effect at the InAs/EuS interface within the detection limit, which is further supported by XMCD measurements. The only detectable proximity effect is a suppression of the Eu moment at the interface layer. Using first-principles calculations, weak magnetization could be induced into InAs layers adjacent to EuS. However, the proximity effect is weak and thus has small influence on InAs. These results are important for the design of topological quantum devices based on hybrid InAs/Al/EuS structures. This study motivates further work on the development of topological quantum devices which operate in the absence of external fields.

## ASSOCIATED CONTENT

**Supporting Information**

The Supporting Information is available free of charge on the ACS Publications website at DOI:

The indexed power spectrum (Fig. S1), the GPA dilatation and rotation maps (Fig. S2), the proposed model (Fig. S3), and the top view based on the model (Fig. S4) of the InAs/EuS interface; The PNR data (Fig. S5 and Table S1); The magnetic measurement and RXRR around Eu $M_{4,5}$-edge (Fig. S6, Fig. S7 and Table S2); The XANES/XMCD at In $L_3$-edge (Fig. S8).

## AUTHOR INFORMATION

**Corresponding Authors**

*E-mail: krogstrup@nbi.dk.

**Notes**

The authors declare no competing financial interests.

## ACKNOWLEDGMENTS

We thank Charles Marcus, Chetan Nayak, Roman Lutchyn, Dmitry Pikulin, Karsten Flensberg, Jaroslav Fabian, Martin Gmitra, Pavel Baláž for helpful discussions and C. B. Sørensen for technical assistance. A.L. acknowledges Ernesto Scoppola for useful discussion and calculations on the EuS absorption scattering length density. Y.L. also would thank Ye Yuan, Chi Xu and Shengqiang Zhou for the support of magnetic measurements, and Jun Deng, Ning Liu and Xiaolong Chen from Institute of Physics, Chinese Academy of Sciences for providing computing resource in magnetism simulation. J.A.K. acknowledges support by the Swiss National Science Foundation (SNF-Grant No. 200021_165910). M.R. acknowledge financial support of the Swiss National Science Foundation project No. CR-SII2 147606. We acknowledge financial support from the Microsoft Quantum initiative, from the Danish Agency for Science and Innovation through DANSCATT, from the European Research Council under the European Union's Horizon 2020 research and innovation program (grant agreement n° 716655), and from the international training network 'INDEED' (grant agreement n° 722176). S.M.S. acknowledges funding from "Programa Internacional de Becas "la Caixa"-Severo Ochoa". S.M.S., C.K. and J.A. acknowledge funding from Generalitat de Catalunya 2017 SGR 327. ICN2 is supported by the Severo Ochoa program from Spanish MINECO (Grant No. SEV-2017-0706) and is funded by the CERCA Programme / Generalitat de Catalunya. Part of the present work has been performed in the framework of Universitat Autònoma de Barcelona Materials Science PhD program. Part of the HAADF-STEM microscopy was conducted in the Laboratorio de Microscopias Avanzadas at Instituto de Nanociencia de Aragon-Universidad de Zaragoza. ICN2 acknowledge support from CSIC Research Platform on Quantum Technologies

Supplementary information

# Coherent Epitaxial Semiconductor - Ferromagnetic Insulator InAs/EuS Interfaces: Band Alignment and Magnetic Structure

Yu Liu[1], Alessandra Luchini[2], Sara Martí-Sánchez[3], Christian Koch[3], Sergej Schuwalow[1], Sabbir A. Khan[1], Tomaš Stankevič[1], Sonia Francoual[4], Jose R. L. Mardegan[4], Jonas A. Krieger[5], Vladimir N. Strocov[5], Jochen Stahn[5], Carlos A. F. Vaz[5], Mahesh Ramakrishnan[5], Urs Staub[5], Kim Lefmann[2], Gabriel Aeppli[5,6], Jordi Arbiol[3,7], Peter Krogstrup[1,*]

[1]Center for Quantum Devices, Niels Bohr Institute, University of Copenhagen & Microsoft Quantum Materials Lab Copenhagen, Denmark.

[2]Niels Bohr Institute, University of Copenhagen, DK-2100 Copenhagen, Denmark.

[3]Catalan Institute of Nanoscience and Nanotechnology (ICN2), CSIC and BIST, Campus UAB, Bellaterra, 08193 Barcelona, Catalonia, Spain.

[4]Deutsches Elektronen-Synchrotron DESY, Hamburg 22603, Germany.

[5]Paul Scherrer Institute, CH-5232 Villigen, Switzerland.

[6]ETH CH- 8093 Zürich and EPFL CH-1015 Lausanne, Switzerland.

[7]ICREA, Pg. Lluís Companys 23, 08010 Barcelona, Catalonia, Spain.

*E-mail: krogstrup@nbi.dk.

# 1. Epitaxial structure of EuS on InAs

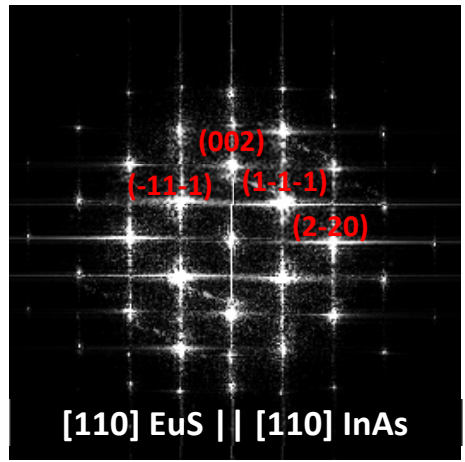

Figure S1 | The indexed power spectrum (fast Fourier transform - FFT) of micrographs in Figure 1 containing both EuS and InAs.

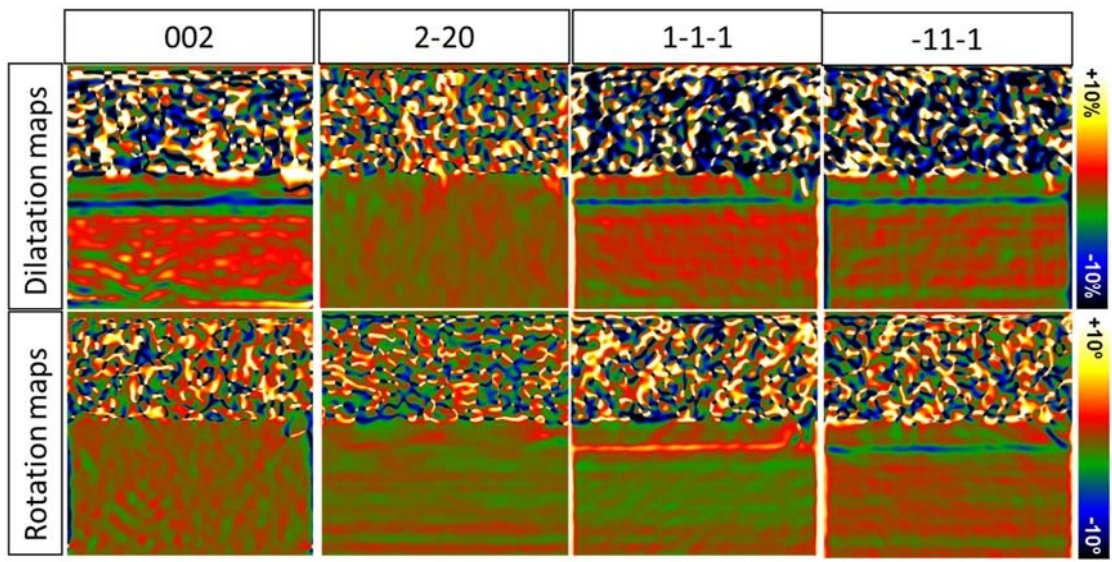

Figure S2 | The GPA dilatation and rotation maps along [002], [2-20], [1-1-1] and [-11-1] of the micrograph displayed in Figure 1a. Note that bulk InAs far away from interface is employed as GPA reference lattice in this work.

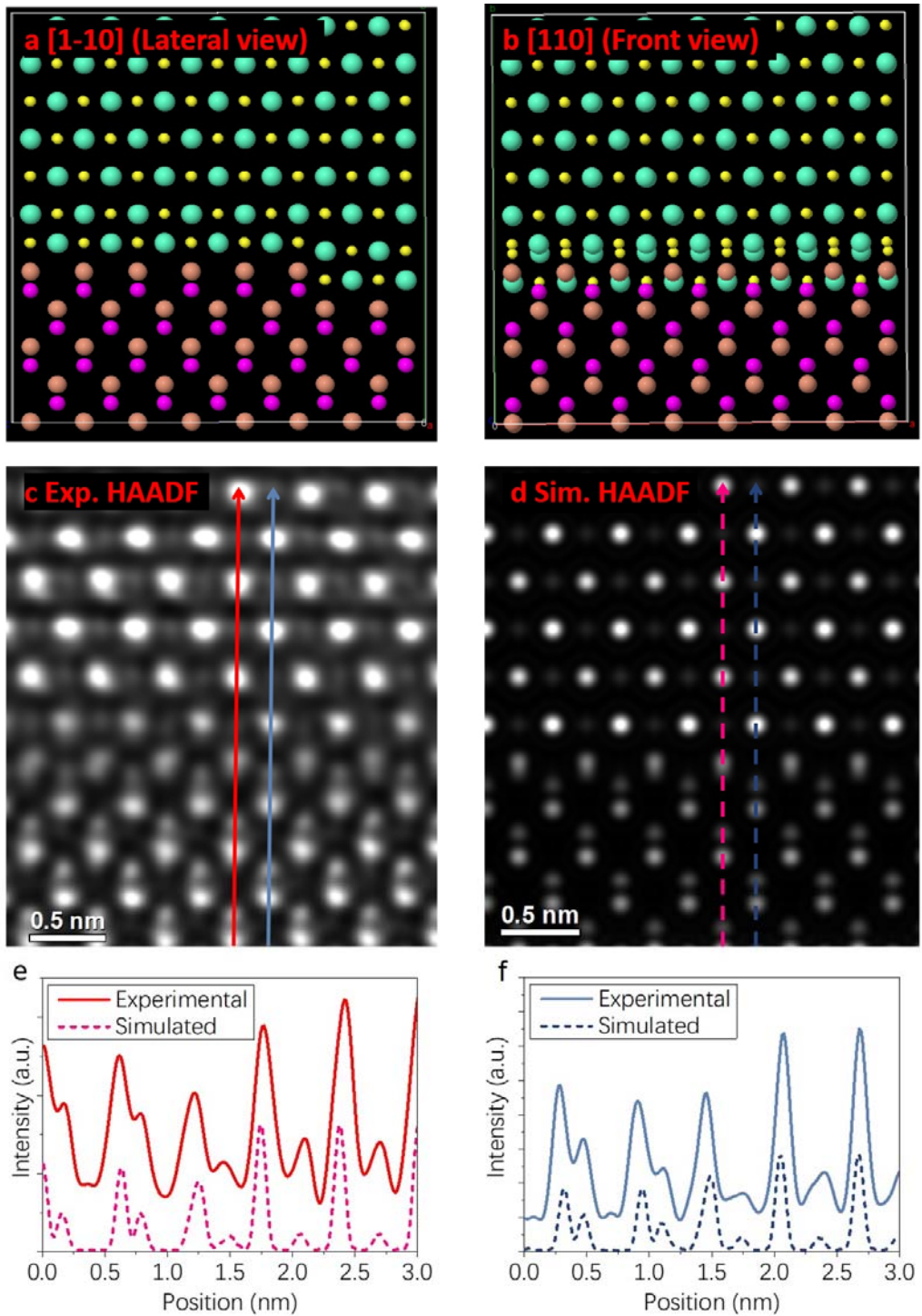

Figure S3 | a,b, The proposed atomic model of InAs/EuS interface projected through [1-10] and [110] directions. c,d, HAADF micrograph and simulated HAADF of the proposed model through the [110] zone axis (front view). e,f, The comparison between the experimental HAADF intensity and the simulated one. The arrows with the corresponding colors or shapes (solid or dashed lines) in c and d show where HAADF intensity curves are extracted.

In the simulation, we used a 300 keV electron beam with a convergence angle of 15 mrad, neglecting aberrations (defocus, astigmatism and higher order aberrations are set to 0). The collection angle goes from 84 to 220 mrad.

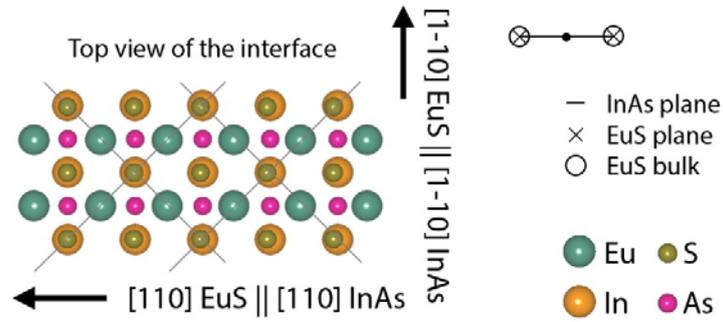

Figure S4 | The top view of the monolayers on the interface, [001] EuS vs. [001] InAs to show lattice match, using the relaxed lattice constants taken from rock salt EuS and zinc blende InAs. Grey lines indicate primitive domains and the parallel and transverse directions are shown by vectors. The PRS is shown on the right panel.

## 2. PNR data analysis

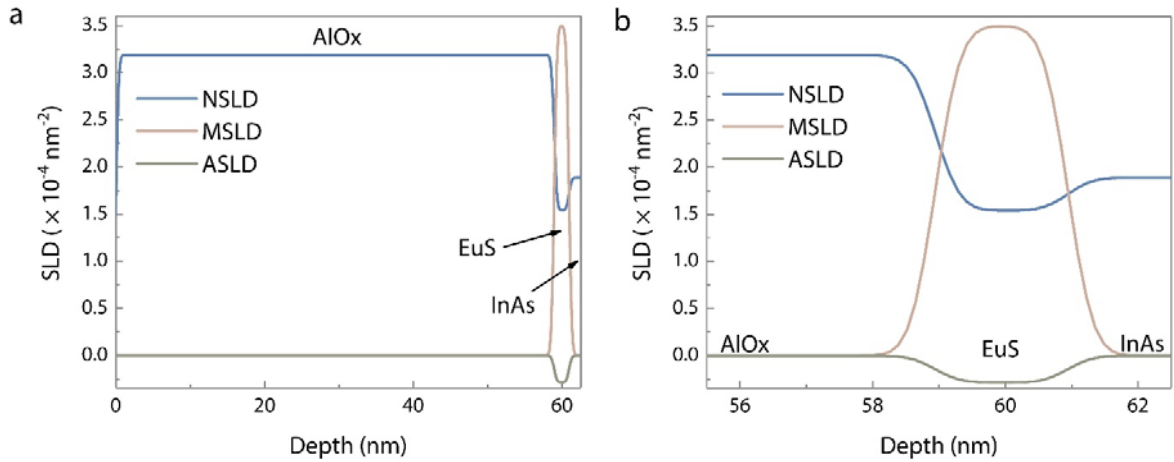

Figure S5 | a, Full-scale depth profiles of PNR nuclear (NSLD), magnetic (MSLD) and absorption (ASLD) scattering length density extracted from 2 K $R^+/R^-$ and 50 K R (the relevant fitting parameters are listed in Table S1). Regarding ASLD, Eu atoms are characterized by a non-neglectable neutron absorption cross-section, which is wavelength-dependent (J. Lynn, et al, Resonance effects in neutron scattering lengths of rare-earth nuclides, Atomic Data and Nuclear Data Tables). This is relevant especially when the experimental data are collected on a time-of-flight neutron reflectometer such as AMOR. However, for the specific structure of our sample, we estimated that a constant absorption scattering length density of $2.4 \times 10^{-5}$ nm$^{-2}$ can still be properly used during data analysis, in agreement with previous results (Katmis, F. *et al.* A high-temperature ferromagnetic topological insulating phase by proximity coupling. *Nature* **533**, 513-516, 2016.). b, The depth profiles are zoomed in to present the magnetic layer.

Table S1 | Structural parameters obtained from PNR data analysis. t = thickness; NSLD = nuclear scattering length density; MSLD = magnetic scattering length density; ASLD = absorption scattering length density; σ = roughness. Parameters reported without error were kept constant during data fitting.

|  | AlOx | EuS | InAs |
|---|---|---|---|
| **t [nm]** | 58.6±0.5 | 2.0±0.2 | bulk |
| **NSLD·10$^{-6}$ [Å$^{-2}$]** | 3.22 ± 0.03 | 1.56 | 1.89 |
| **MSLD·10$^{-6}$ [Å$^{-2}$]** | 0 | 3.50±0.08 | 0 |
| **ASLD·10$^{-6}$ [Å$^{-2}$]** | 0 | 0.27 | 0.01 |
| **σ [Å]** | 3±1 | 3±1 | 3±1 |

## 3. RXRR measurement

RXRR measurements were carried out at the RESOXS end station present at the SIM beamline of the Swiss Light Source (SLS) at the Paul Scherrer Institute (PSI), Switzerland, in order to determine the magnetic profile structure of the EuS layer by probing the Eu $M_{4,5}$ edges, which correspond to atomic transitions from the 3d core levels to empty states in the 4f orbitals. The nominal sample structure for the RXRR measurements is As (20 nm)/EuS (2.5 nm)/InAs (15 nm)/GaSb (bulk). From SQUID measurements (Fig. S6, carried out in a MPMS Quantum Design system at the Laboratory for Mesoscopic Systems, ETH Zurich, Switzerland, and the Laboratory for Multiscale Materials Experiments, Paul Scherrer Institute, Switzerland) we find a critical temperature of 23 K and a bulk-like magnetic moment (6.0 $\mu_B$/Eu at 4 K, or ~6.5 $\mu_B$/Eu extrapolated to 0 K). The sample was positioned at the end of a cold finger, whose base temperature (13 K) was estimated by following the variation of the magnetic asymmetry at the Eu $M_5$ edge as a function of temperature. The sample was cooled down from room temperature to the base temperature in the presence of a magnetic field (~0.07 T) generated by a set of permanent magnets to saturate the magnetization in the sample plane. This sample plane is also the scattering plane. Reflectivity measurements at fixed energy and energy scans at a fixed momentum transfer value were carried out with both left and right polarized light. In addition, a reflectivity scan above the critical temperature was also taken to extract the sample structural parameters without the magnetic component. Given the strong dependence of the scattering factors near the resonance and the strong absorption at these photon energies (1131.5 eV for the $M_5$ edge shown in Fig. S7), RXRR measurements constitute a very sensitive probe of the electronic and magnetic structure at InAs/EuS interfaces.

To analyze the data, we calculated the scattering factors of EuS based on the magnetic x-ray absorption data for EuO published in the literature [Holroyd et al JAP 95 (2004) 6571] and the Kramers-Kronig transformation. This is expected to be sufficiently accurate, given that the 4f shell participates only weakly in chemical bonding, such that its spectral and magnetic characteristics remain identical for similar rock salt compounds. On the other hand, possible birefringence effects were not taken into account. In fitting the data, we assumed also the simplest sample structure that could fit the data reasonably (for example, we did not consider a modified As surface layer due to air exposure). Also, in fitting the data, which was carried out with the Dyna software [Elzo et al, Journal of Magnetism and Magnetic Materials 324 (2012) 105], both the reflectivity and energy scans for both light polarizations were used; we started with the data at high temperature (29 K) to extract the structural parameters, which were used as starting values for the fits of the low temperature data.

The reflectivity data shows the presence of a long frequency modulation, due to the thin EuS layer, accompanied by a high frequency modulation due to the thick layers in the sample. From the fitting to the data, we find that the interfaces are atomically sharp, with interface roughness ranging from 2-6 Å. It is critical that the beatings of the low frequency signal do not match for the left and right circularly polarized light, indicating an asymmetry in the magnetic scattering; in other words, a uniform magnetic layer cannot account for such an asymmetry in the data. To allow the presence of an asymmetry in the magnetic layer, we considered the presence of an interfacial layer, on both the InAs and As interfaces. The result of the fitting shows that the thickness of the EuS layer at the As interface layer reduced to zero, while that at the InAs layer remained finite, with a strongly modified magnetic moment value. The best fitting to the data is shown in Fig. S7 and was obtained for a 1.4 Å thick interface layer with a magnetic moment reduced by a factor of 2.7 from the bulk value.

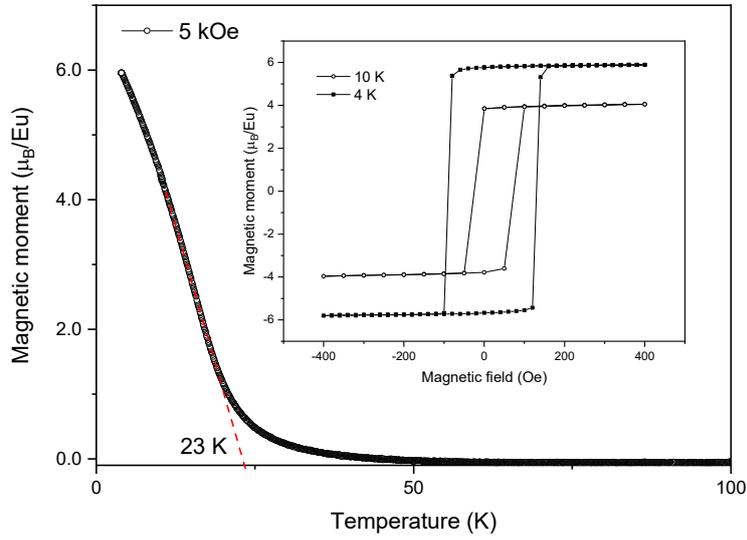

Figure S6 | Magnetic moments varying with temperatures of the sample used for RXRR from 5 to 100 K under a field of 5 kOe. Inset: The magnetization as a function of magnetic field within a 400-Oe magnetic field at 4 K and 10 K for the same sample.

Table S2 | Structural parameters obtained from RXRR data analysis. t = thickness; σ = roughness; XRSLD = X-ray scattering length density; MMS = magnetic moment $\mu_B$/Eu (value calibrated to the SQUID data at 13 K).

|  | As | EuS | EuS at the InAs/EuS side | InAs | GaSb |
|---|---|---|---|---|---|
| t [Å] | 864.6 | 20.5 | 1.4 | 176.5 | bulk |
| σ [Å] | 7.5 | 2.5 | 1.9 | 1.8 | 6 |
| XRSLD [Å$^{-2}$] | 0.07635 | 0.03125 | 0.03125 | 0.02987 | 0.02932 |
| MMS | 0 | 3.4 | 1.3 | 0 | 0 |

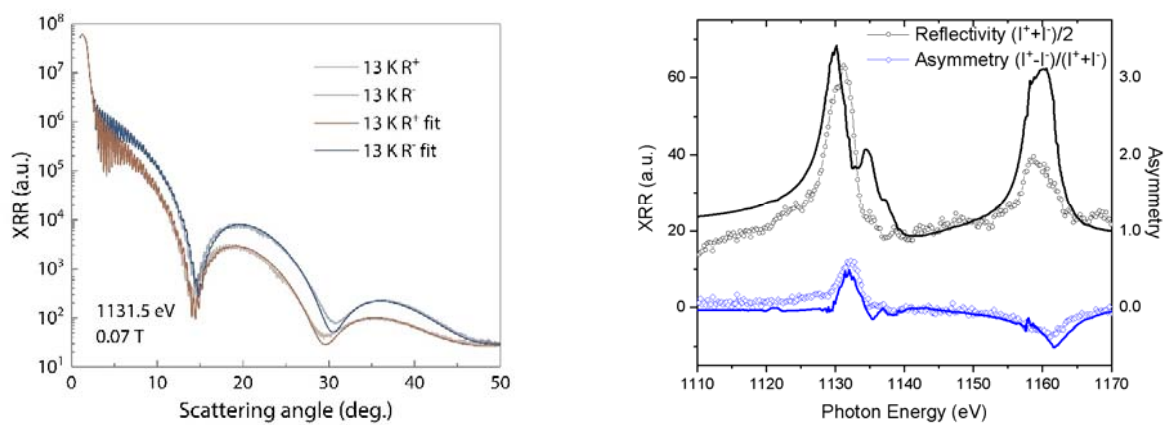

Figure S7 | Left panel: Experimental data of RXRR collected under 0.07 T at 13 K together with the respective fitting curves. The reflectivity curve from the sample was measured as a function of the scattering angle using a photon energy of 1131.5 eV (at the maximum of the asymmetry signal indicated in the right panel). The relevant fitting parameters are listed in Table S2. Right panel: Polarization-averaged and asymmetry reflectivity energy scans collected at the Eu $M_{4,5}$-edge under 0.07 T measured at 13 K and at a constant momentum transfer value of 0.0894 Å$^{-1}$.

## 4. XANES and XMCD data from In

XANES/XMCD measurements were carried out as a function of temperature at the In L3 absorption edges. As shown in bottom panel of Fig. S8, the overlapping between the XMCD spectra obtained at 5 K and 20 K suggests that the magnetic signal around 3740 eV is an artefact.

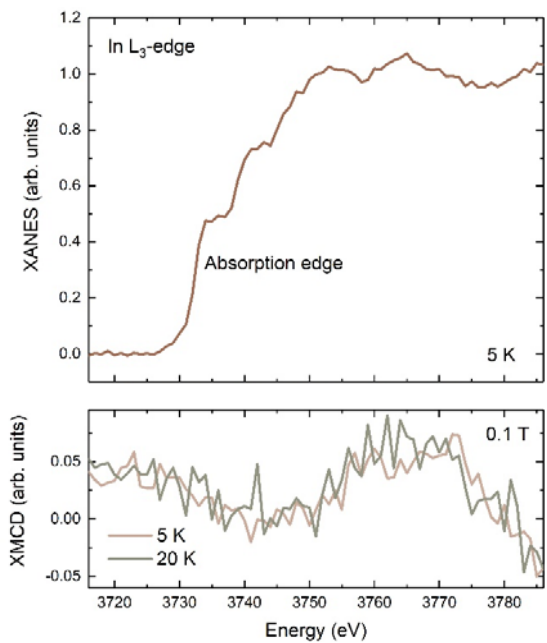

Figure S8 | Top panel: XANES spectrum collected at the In $L_3$-edge at 5 K; Bottom panel: XMCD spectra at the In $L_3$-edge under 0.1 T measured below (5 K) and above (20 K) the EuS Curie temperature.